\documentclass[a4paper,11pt]{article}
\pdfoutput=1
\usepackage[utf8x]{inputenc}
\usepackage{amsmath,amssymb,mathtools}
\usepackage{pdfpages} 
\usepackage{tikz}
\usepackage{hyperref}
\usepackage{xcolor}
\usepackage{multirow}
\usetikzlibrary{3d,calc}
\usepackage{graphicx}
 \graphicspath{ {LaTeXDocs/}}
 \usepackage[toc,page]{appendix}
 \usepackage{tikz}
\usetikzlibrary{trees,arrows}
\tikzset{level 1/.style={level distance=1.5cm, sibling distance=3.5cm}}
\tikzset{level 2/.style={level distance=1.5cm, sibling distance=2cm}}

\usepackage{hyperref}

\def\beq{\begin{equation}}
\def\bee{\begin{equation}}
\def\eeq{\end{equation}}
\def\bea{\begin{eqnarray}}
\def\eea{\end{eqnarray}}
\def\bd{\begin{displaymath}}
\def\ed{\end{displaymath}}

\usepackage[T1]{fontenc}

\tikzset{bag/.style={text width=30em, text centered,yshift=-0.2cm}}
 \usepackage[text={6in,8in},centering]{geometry}
 \usepackage{pdfpages}
\title{ Massive   AdS Supergravitons   and Holography } 
\author{}

\date{ }

\begin{document}

\maketitle
\begin{center}
Constantin Bachas \\
\medskip\medskip\medskip

\textit{Laboratoire de Physique  de l'\'Ecole Normale Sup\'eri{e}ure 
 \footnote{\,\'Ecole Normale Sup\'erieure, Université PSL, Paris, France}\ \footnote{\,Centre National 
 pour la Recherche Scientifique}\ \footnote{\,Sorbonne Université} 
\,\footnote{\,Université Paris-Diderot,
Sorbonne Paris Cité, Paris, France} \\ 
24 rue Lhomond, 75231 Paris Cedex, France
}

\vskip 2cm 
 
\begin{abstract}

I compare two holographic mechanisms giving to the 
 graviton  a parametrically-small  supersymmetric mass $m_g$  in Anti-de Sitter  spacetime. 
In the context of bimetric gravity these mechanisms couple  `weakly'  two initially  decoupled 
superconformal theories by:  (i) turning on a double-trace deformation, or (ii) gauging a common global
symmetry.  Superconformal invariance  restricts   the number of Poincar\'e supercharges to 
   $N_Q\leq 4$  for mechanism (i) and  to
    $N_Q\leq 8$ for mechanism (ii), and the  AdS dimension to 
$D\leq 5$.  The\\  \vskip -4.2 mm\hskip 1.5mm  putative effective supergravities are expected to 
break down in the   $m_g\to 0$ limit  
at an intermediate scale between   $ m_g$ and  $m_{\rm Planck}$. 
 In  a recently-proposed   microscopic embedding of  
    mechanism (ii)  in string theory \cite{Bachas:2017rch, Bachas:2018zmb},  
    I show that $m_g=0$ is at infinite distance in moduli space, and I
  relate the breakdown of the effective theory  to the condensation of 
  unprotected spin-2 excitations
  in  the string-theoretic description  of the  `holographic bridge'.  
  The phenomenon is invisible  in  the  weakly-coupled CFT side.  
    I  conjecture that similar phenomena should be  at work 
   in all cases.

\hfill \break
\hfill \break
 
\end{abstract}

\end{center}

\newpage
\newpage

\setcounter{footnote}{0}

\section{Introduction}

There has been considerable  interest in recent years  in  relativistic theories of massive gravity, see  e.g.\,\cite{Hinterbichler:2011tt}  for  reviews.
The  model-independent  upper bound on the graviton mass\,\footnote{This 
bound comes from   the  dispersion relation. For other   bounds on $m_g$ see 
\cite{deRham:2016nuf, Will:2018gku}.}
from the  LIGO/Virgo collaboration   \cite{Abbott:2017vtc}
 is  $m_g \lesssim  7.7\times 10^{-23}$ eV,   which  translates to  a  lower bound on its
Compton wavelength 
 $\lambda_g \gtrsim \,  1.5 \times 10^{13} $km  (or about 1.5 light years). It  leaves 
ample room for   a tiny but finite graviton mass  with   dramatic implications
for the large-scale structure of the  Universe. 

       An important question  is whether effective low-energy theories of  massive gravity 
       have a consistent ultraviolet completion, or whether they  are in  the 
       quantum-gravity `swampland'.  A  direct embedding of
        massive  Anti-de Sitter  (AdS) gravity
        in string theory was recently proposed in refs.\,\cite{Bachas:2017rch,Bachas:2018zmb}. 
        Besides living in AdS$_4$,   this embedding has ${\cal N}=4$ supersymmetries  and is
        clearly not realistic.   It does,  however,  allow to address  the breakdown
        of the effective massive-gravity  theory in a controlled setup.  Indeed, a common  
        feature of all effective theories of massive gravity
        is that their
       range of validity vanishes when  $m_g \to 0$, as  follows by inspection of the non-linear
       action of the Stueckelberg  field \cite{ArkaniHamed:2002sp}. 
         In  
        AdS$_4$  the breakdown  scale is {  at most}
         $\Lambda_* \sim  (m_g m_{\rm Pl}/l_{\rm AdS})^{1/3}$, where
         $l_{\rm AdS}$ is the AdS radius and   $m_{\rm Pl}$ is  the 
         four-dimensional Planck scale 
            \cite{deRham:2016plk}.  In this letter I will    identify   the  origin  of the  breakdown  in the 
            string-theory embedding of  \cite{Bachas:2017rch, Bachas:2018zmb}.

            A  second question that I will address is whether similar  string theory embeddings exist in other
            dimensions and with different amounts of supersymmetry. It is conceptually simpler 
            to consider the more general   bimetric theories,  from which massive gravity can be obtained as
            a limit \cite{Hinterbichler:2011tt}. Assuming  that 
           all AdS   vacua have   holographic duals, one is led to study the following setup: 
              two initially decoupled conformal  theories  are made to interact `weakly' in a sense to
              be made more precise below. After coupling,  the two conserved energy-momentum tensors
              mix so that their  sum remains conserved while  an orthogonal combination acquires a
              small anomalous dimension $\epsilon_g$. This is the CFT description of   two decoupled metrics
              interacting  weakly,  so that one combination obtains a mass while the  second stays massless.
                  The interaction of the  conformal theories can  be one of two kinds:
     \smallskip
     
     $\bullet$ A  `double trace'  deformation $\int \hskip -1mm \phi \Phi$,  with  $\phi$ and 
     $\Phi$   operators of the separate theories;  \smallskip
  
 $\bullet$ Coupling through a messenger field, such as  the  gauging of a common global symmetry.
       \vskip 1mm
    \noindent  Note that the second mechanism  differs   from the first since
      massless messengers cannot be integrated out to give 
    local double-trace interactions. 
    
     \smallskip
      These   mechanisms are  constrained by the requirement   of superconformal symmetry $(\mathfrak{g})$. 
      The massless supergraviton  is dual to a conserved energy-momentum tensor that belongs
      to a short representation of  $\mathfrak{g}$.  Higgsing  combines it with a Stueckelberg
      supermultiplet into a long representation with  anomalous scaling  dimension  $ \epsilon_g $,  where
        $  \epsilon_g  \sim  m_g^2 $  is parametrically small.  The   list of all unitary
      superconformal representations is  given  in ref.\,\cite{Cordova:2016emh}. 
      One sees by inspection   that for SCFTs  in more than four  dimensions,  or  with 
       more  than half-maximal supersymmetry,    the massless
       spin-2 multiplets  are absolutely protected, i.e. they cannot  combine to form long representations.
       The corresponding
      massive supergravities are thus  a priori  ruled out.  The double-trace mechanism is further
      constrained by the requirement that the  bridging  operator be  of product form,
      which is impossible with more  than ${1\over 4}$-maximal supersymmetry. These bounds are
      saturated,  as one can show with examples. Furthermore in all the allowed cases  the 
     Stueckelberg multiplets are  made out of the ingredients of the `bridge', namely (i) the
      operators $\phi$ and $\Phi$ ,  or (ii)   currents and free vector multiplets. This is a
      consistency check for these mechanisms.

         Double-trace CFT deformations are  described   by modified 
         boundary conditions in AdS$_D$  \cite{Witten:2001ua, Berkooz:2002ug}, but their 
           lift  to   string theory  (or indeed simply  to  ten dimensions)   is problematic. 
     They were studied  as a  mechanism for Higgsing gravity in
      refs.\,\cite{Kiritsis:2006hy, Aharony:2006hz, KN},  following up on  the original proposal 
      by Porrati \cite{Porrati:2001db}  in the single-metric limit. The gauging mechanism 
      in \cite{Bachas:2017rch, Bachas:2018zmb}  shares many features  of  [the AdS version of]
      Randall-Sundrum models \cite{Randall:1999ee, Karch:2000ct}.  
      As the first  microscopic implementation of these ideas on the gravity side, 
       it will allow us to study the breakdown of the putative 
      effective theory. I will argue that the full string-theoretic description
      of the bridge (in terms of  a cutoff   AdS$_5$ Janus throat)  is essential for  understanding  
       the $m_g\to 0$ limit. 
       
       An outstanding question  is whether similar microscopic resolutions
        exist for  all  the other  cases of holographic graviton Higgsing. 
       Another interesting problem is the construction of   massive supergravities,
       in particular  half-maximal ones  in AdS$_4$  and AdS$_5$.\,\footnote{A 
       ${\cal N}=4$  supergravity with four-derivative terms and 
    a massive spin-2 mode has been
    constructed in  ref.\,\cite{Ferrara:2018wlb}. 
    Here I refer however to standard two-derivative actions.}  
       These theories  should be highly constrained and could be compared with the 
          ghost-free bosonic  action  of de Rham 
          et al \cite{deRham:2010ik, Hassan:2011vm}.\,\footnote{There have been
          lingering concerns about   pathologies of  massive-gravity actions, see e.g. 
           refs.\,\cite{Chamseddine:2013lid, Deser:2014fta}. These arise, in essence,   because
           one tries  to make two disjoint metrics interact via an effective  local 
           action. 
           The microscopic description of the  interaction given in  \cite{Bachas:2017rch, Bachas:2018zmb} 
            supports the view
            that, at least 
          perturbatively  around  AdS spacetime, these concerns can be laid to rest.
          }
                I hope to return to these questions in the near future.

  \smallskip  
   The plan of this paper is as follows. In section \ref{sec:2}\,  I explain  why 
     massive Anti-de Sitter  supergravities  with more than half-maximal
   supersymmetry, as well as  the cases ${\cal N}=1$ in AdS$_6$ and AdS$_7$, are ruled out. 
   In section
   \ref{sec:3} I describe the  two holographic mechanisms for the superconformal  Higgsing 
   of the  graviton, and argue  that only ${1\over 4}$-maximal supersymmetry is possible
   in the double-trace case. Supersymmetry plays a minor role in most of the literature on
   massive gravity. But by  guaranteeing  the stability of AdS vacua and of their weak Higgsing
   interaction, it is bound to be  of great help in any effort to embedd massive gravity
   in  string-theory.

   The reader not interested in the supersymmetric  details  
   can skip  altogether  subsections    \ref{sec:3.2} and \ref{sec:3.3}.     
      Section \ref{sec:4} contains the main message of the paper and can be read independently. There I 
       trace  the  breakdown of the effective theory
    when $m_g\to 0$ to  towers of spin-2 modes in the string-theoretic
    description  of  the  Higgsing  mechanism   proposed    in
     \cite{Bachas:2017rch}\cite{Bachas:2018zmb}. 
    The  bridge is in this case a cutoff AdS$_5$ Janus throat connecting two AdS$_4$ vacua.
    Although the background supersymmetry is important for stabilizing the entire setup, the
    condensing spin-2 modes are unprotected and only visible on the string-theory side.
    I conjecture that  similar phenomena  might  explain the breakdown in all other
    cases as well.

    
  \section{No go for $D>5$  and  for $N_Q>8$}\label{sec:2}  
    
  We are interested in   AdS$_D$  solutions of string theory  around which the metric fluctuations
   have a mass,  and  the dimensionless 
    parameter $m_g l_{\rm AdS}$ can be tuned arbitrarily close  to zero. 
     Here $l_{\rm AdS} $ is the radius of  AdS$_D$,    and $m_g$ is 
   the   graviton  mass defined so that for  the  pure Einstein theory with cosmological constant 
     $m_g=0$. 
    The mass  need not be a continuous  
    parameter, all 
    we   assume is  that it can be made  parametrically  small. 
  
     Field  excitations in   AdS$_D$   are in  unitary 
   representations of the  conformal algebra $\mathfrak{so}(2,D)$.  
  If  the solution is supersymmetric the field supermultiplets are assigned to
   representations of the  relevant superconformal algebra  $\mathfrak{g}\supset \mathfrak{so}(2,D)$.
    A  massless spin-2 particle  is in a short representation of the algebra.  To obtain  mass 
    it must  combine
     with  a  representation that provides the missing polarizations of the
    massive multiplet.  The resulting representation is long, and its mass is unrestricted.
     Schematically 
 \bea
 {\rm [Massive/Long]} \ 
 \xrightarrow[m_g \to 0 ]{ }  \  {\rm [Massless/Short]}\, \oplus\, {\rm [Stueckelberg]}\ . 
 \eea
   Now  for many superalgebras  the massless supergraviton is {\emph {absolutely protected}}, 
   i.e. it never appears  in the decomposition of a long multiplet.
    When this is the case the graviton cannot obtain a mass without breaking some of  the supersymmetries.   In such cases  the existence  of  AdS supergravity  
    with tunable   graviton mass
    can be  excluded.
  
  \smallskip
      I    restrict  the discussion  to the range $4\leq   D  \leq 7$, or equivalently to the range
      $3\leq d\leq 6$ where $d$ is the dimension of the dual SCFT.
        There are  no superconformal 
  theories  in  higher dimensions   \cite{Nahm:1977tg},
  while in  $D<4$ there is no dynamical  graviton.\,\footnote{The case of  AdS$_3$  is nevertheless
  very interesting, for   massive $D=3$ supergravities
  see e.g.  refs.\,\cite{Bergshoeff:2009hq, Andringa:2009yc}. 
 Three dimensions is   special for many reasons, in particular because   the conformal group 
  is infinitely extended. }
      All   unitary superconformal representations   in  this range have been listed 
   in ref.\,\cite{Cordova:2016emh}.  The cases in  which  
    the massless supergraviton is absolutely protected  are given  in 
   table \ref{tab:11}\,.  I have adopted here the unifying notation of \cite{Cordova:2016emh}
   in which a representation is denoted by the Dynkin labels of  the superconformal primary,
   the letters $A$  or $B, C\cdots $  indicate whether a short  representation is at the unitarity threshold
   or separated from the continuum by a gap, and the subscript gives the level
   of the first null states. Since the reader might not be familiar with this notation, I will
   translate to more standard ones where appropriate.

   \begin{table}[!htbp]
   \vskip 5mm
 \begin{center}
\begin{tabular}{   c |c | c |c |}
 \cline{2-4}   
 &  {\bf Susy} & \bf $\mathfrak{g}$  & \bf \,    Massless  graviton  \\   [0.7ex]    \hline\hline 
  \multicolumn{1}{ |c |} {\multirow{2}{*}{\,\,\, AdS$_7$}  \,\,\, }  & 
\multicolumn{1}{ | c| } {${\cal N}$=(2,0)} & $\mathfrak{osp}(8^*\vert 4)$  & $D_1[0,0,0]_4^{(0,2)}$ \\    [0.5ex]  \cline{2-4}  
\multicolumn{1}{ |c| }{}    &
  \multicolumn{1}{ |c |} {${\cal N}$=(1,0)}   & $\mathfrak{osp}(8^*\vert 2)$ & $B_3[0,0,0]_4^{(0)}$ \\      [0.5ex]   \hline\hline  
 \multicolumn{1}{ |c |}    {AdS$_6$}   & ${\cal N}=1$   & $\mathfrak{f}(4)$ & $B_2[0,0]_3^{(0)}$   \\      [0.5ex]   \hline\hline
  \multicolumn{1}{ |c |} {\multirow{2}{*}{\,\,\, AdS$_5$}  \,\,\, }  & 
\multicolumn{1}{ | c| } {${\cal N}=4$} & $\mathfrak{psu}(2,2\vert 4)$ & $B_1\bar B_1[0;0]_2^{(2,0,2)}$ \\    [0.5ex]  \cline{2-4}  
\multicolumn{1}{ |c| }{}    &
  \multicolumn{1}{ |c |} {${\cal N}=3$}   & $\mathfrak{su}(2,2\vert 3)$ & $B_1\bar B_1[0;0]_2^{(1,1;0)}$ \\      [0.5ex]   \hline\hline
   \multicolumn{1}{ |c |} {\multirow{4}{*}{\,\,\, AdS$_4$}  \,\,\, }  & 
\multicolumn{1}{ | c| } {${\cal N}=8$} & $\mathfrak{osp}(8\vert 4)$  & $B_1[0]_1^{(0,0,0,2)\ {\rm or}\ (0,0,2,0)}$ \\    [0.5ex]  \cline{2-4} 
\multicolumn{1}{ |c| }{}    &
  \multicolumn{1}{ |c |} {${\cal N}=7$}   & $\mathfrak{osp}(7\vert 4)$ & $B_1[0]_1^{(0,0,2)}$ \\      [0.5ex] \cline{2-4}
  \multicolumn{1}{ |c| }{}    &
  \multicolumn{1}{ |c |} {${\cal N}=6$}   & $\mathfrak{osp}(6\vert 4)$ & $B_1[0]_1^{(0,1,1)}$ \\      [0.5ex] \cline{2-4}
  \multicolumn{1}{ |c| }{}    &
  \multicolumn{1}{ |c |} {${\cal N}=5$}   & $\mathfrak{osp}(5\vert 4)$ & $B_1[0]_1^{(1,0)}$ \\      [0.5ex] \hline\hline
 \end{tabular}  
   \medskip  
\caption{\scriptsize  The  AdS$_D$ supergravities for which the massless  graviton   is 
 in an   
absolutely protected representation. We list the number of supersymmetries, the superconformal algebra
$\mathfrak{g}$, 
and the absolutely protected 
 representation   in the 
  notation of  ref.\,\cite{Cordova:2016emh}. 
 The list  includes  all  cases in $D>5$ dimensions,  
and all cases with
more than half-maximal supersymmetry  for reasons explained  in the main  text.  
The table  has  
 redundancies:   when   the massless graviton is   protected for   given $(D, {\cal N}_0)$ 
  it is also protected  for all  $(D, {\cal N} >{\cal N}_0)$.  
   }   \label{tab:11} 
 \end{center}
 \end{table}
   
     The multiplet of the 
   massless AdS  graviton, or of the dual   
   conserved   energy-momentum tensor, is always  in  a short representation of  $\mathfrak{g}$.
   This is an  $A$-type representation 
    when  Higgsing is allowed
        (see
     section \ref{sec:3}) and a gapped,  $B$- or $D$-type,  
   representation in all excluded cases. 
     Note that the gap does  {\it not}  guarantee  absolute  protection of a generic multiplet. 
   It is a necessary, not a sufficient condition. 
   
   \smallskip

     The forbidden list of table \ref{tab:11} 
       includes all cases with more  than half-maximal supersymmetry,
      i.e. with  $N_Q> 8$ Poincar\'e supercharges, as advertized. 
        The half-maximal cases
   in $D=6,7$ dimensions are also excluded.\,\footnote{\,For $D=6$ 
(i.e. for SCFT$_5$)  maximal 
   supersymmetry is not compatible  with conformal
   invariance.}   
This  ties  in nicely with the fact that 
 holographic mechanisms for Higgsing the supergraviton  are not available
for  these   $(D, {\cal N})$ pairs, as will be clear in the following section. 
A quick mnemonic for string theorists is that Higgsing is only possible if  the dual SCFT$_d$ 
can couple as a defect to  a higher-dimensional bulk  theory without breaking   its own 
superconformal symmetries.  
This is indeed impossible whenever $N_Q>8$ or for $d=5,6$.\,\footnote{Our counting of supersymmetries
in the table is the standard counting on the field theory side,
 i.e. ${\cal N}$ is the number of spinor supercharges
of the SCFT$_d$.  Superconformal symmetries double this number on the gravity side. The two
maximal massive supergravities in particular,  dual to
${\cal N}=4$ SCFT$_3$ and ${\cal N}=2$ SCFT$_4$,  correspond
to  AdS$_D$  vacua  of  gauged $N=4$  supergravity in $D=4$ and $D=5$ dimensions.  In both
cases the number of Poincar\'e supercharges of the SCFT  is $N_Q=8$.}

    Let me  stress  however  that  these   no go statements 
  are  purely kinematical 
     and do   not depend on  any  details of the   Brout-Englert-Higgs (BEH)   mechanism. The only assumptions 
   are  superconformal invariance and unitarity.

   It should be also noted that the existence of   marginal superconformal deformations is neither
  a necessary nor  a sufficient condition for the Higgsing of the supergraviton. Indeed 
   superconformal manifolds exist for  ${\cal N}=1,2, 4$ in $d=4$ but  only for 
     ${\cal N} \leq 2$ in $d=3$    
      \cite{Cordova:2016xhm}.  On the other hand, 
        Higgsing of the supergraviton   is forbidden   for  ${\cal N}=4$ in 
      AdS$_5$ whereas 
        it is possible    for
          ${\cal N}=4$ in 
      AdS$_4$. 
         

\section{Holographic BEH  mechanisms}\label{sec:3}

     Let us  take now a closer look at  the mechanism  by which the graviton obtains a mass. 
   The most convenient starting point is a   bimetric   theory   with  effective action 
        \bea\label{2}
        S =  {m^{D-2}_{\rm Pl}\over 2}\int  \sqrt{g}\, (R[g] + {\cal L}_{\rm m} )  + 
         {M^{D-2}_{\rm Pl}\over  2}\int  \sqrt{G}\, (R[G] +  {\cal L}_{\rm m}^\prime ) \, + \, S_{\rm int}(g, G)\ , 
        \eea
   where $g$ and $G$ are two independent metrics. Each of  them has
    its own Einstein action with  negative cosmological constant,   and they couple   minimally to  separate
  matter fields
  whose Lagrangians are  ${\cal L}_{\rm m}$ and ${\cal L}_{\rm m}^\prime$. 
  In the absence of the interaction, $S$  describes two decoupled Universes. We assume
  that they both have  
 CFT  duals and work at leading order in $S_{\rm int}$.  At this order,   $S_{\rm int}$
 does not affect  the classical backgrounds $g_0$ and $G_0$,   but it  mixes the two metric  fluctuations $g-g_0$
 and $G-G_0$ making one linear combination massive.  The limit $M_{\rm Pl}\to\infty$
 decouples  the massless mode leaving  a theory of a single massive graviton. 
   
     Note that in order to  write  an interaction term that preserves both reparametrization symmetries
     one needs  a diffeomorphism  between the two Universes, $X^M(x^\mu)$. This is  the Stueckelberg
     field  that provides the missing polarizations of the massive graviton.

 
   \subsection{Gauging versus double-trace deformations}

     The holographic viewpoint  of this setup  is  as follows. The non-interacting Universes
     are dual to   decoupled conformal field theories,
     cft$_d$  and CFT$_d$, whose  energy-momentum   tensors,
     $t_{ab}$ and $T_{ab}$,   are  separately conserved -- they  are dual to the two massless
   gravitons. The   2-point function of $t_{ab}$  [normalized by  canonical Ward identities]  reads
       \bea
 \langle t_{ab} (x) t_{ce}(0)\rangle = {c/2\over x^{2d}} \left( I_{ac}(x)
 I_{be}(x) +  I_{ae}(x)
 I_{bc}(x) - {2\over d} \eta_{ab}\eta_{ce}\right) \ , 
 \eea 
   where   $ I_{ab}(x) = \eta_{ba} - 2x_a x_b/x^2$   and  the central charge $c$ is  related to the 
  radius of  the Anti-de Sitter  background (whose  metric is $g_0$)      by the relation
   \cite{Aharony:2006hz}
   \bea
    c = { (m_{\rm Pl}\, l_{\rm AdS})^{d-1} } \frac{d(d+1)\Gamma(d)}{ 2\pi^{d/2}(d-1)\Gamma ({d\over 2})} 
              \ \  .         \eea           
   Similar expressions hold 
   for   $t\to T$,   $c\to C$,  $m_{\rm Pl} \to M_{\rm Pl}$ and $l_{\rm AdS}$ replaced by $L_{\rm AdS}$, 
   the radius of the  second AdS metric $G_0$.

   Before turning on an  interaction, both energy-momentum tensors have canonical scaling dimension
   $[t]= [T] = d$. The interaction  splits  this degeneracy.   
    This is a problem of second-order degenerate perturbation theory,  with the extra input that
    since the total energy-momentum tensor is still conserved   its scaling dimension is unchanged. 
   This determines at leading order in the perturbing Hamiltonian 
   the two orthonormal energy eigenstates. They correspond (by the operator-state correspondence) to 
   the two   spin-2  operators
   \bea\label{2spin2}
    {\cal T}_{ab} =  {{1\over \sqrt{c +C }}}  (t_{ab}+T_{ab}) + O(\lambda)   \quad {\rm and} \quad
     \tilde{\cal T}_{ab} =  {1\over \sqrt{Cc^2+cC^2}}  (C t_{ab}- c T_{ab}) + O(\lambda) \ ,  
   \eea
     where $\lambda$ is 
     the small parameter of the perturbation.  Clearly, the conserved energy-momentum tensor
     has canonical dimension $[{\cal T}] = d$ and is dual to a massless graviton, whereas 
     the other,  
     orthogonal  combination acquires an anomalous dimension $[\tilde {\cal T}]   = d + \epsilon_g$ and is  
     dual in AdS  to a massive spin-2 particle.\,\footnote{The relation between scaling dimension and mass
     for spin-2 states is $m_g^2 \,l_{\rm AdS}^2 = \Delta (\Delta -d)$.  In warped compactifications
     both the AdS radius and the mass vary and only  their product,  which  stays fixed,  has invariant
     meaning. For the case at hand if the two AdS radii are different we have 
     $m_g^2 \,l_{\rm AdS}^2 = M_g^{ 2}  \,L_{\rm AdS}^{ 2}$. 
     }

                For  CFTs coupled by a marginal
                 double-trace operator,   $\epsilon_g$  
                 was computed  in ref.\, \cite{Aharony:2006hz} with the following   leading-order result
                 \bea\label{ahetal}
                 \epsilon_g  =    \lambda^2 ({1\over c} + {1\over C})   + O(\lambda^3) \ . 
                 \eea
      Numerical factors that depend on                                                                                                                                                               the perturbing operator and the  dimension $d$ (but not on the central charges) have been  absorbed in the
   coupling  $\lambda$.  One  sees  that the decoupling limit $M_{\rm Pl}\to\infty$,  which freezes the fluctuations of the auxiliary metric $G$ in massive-gravity theories,  amounts to sending 
       $C\to\infty$  and hence  ${\cal T}_{ab} \to 0$ and
      $\tilde {\cal T}_{ab}  \to  {{1\over\sqrt{c}}\, t_{ab}}$. This  leaves   a single  massive spin-2 operator
      as expected.    
 

    So  a bimetric  theory is dual in holography 
    to  two conformal field theories that are  weakly coupled to
    each other.  But what does `weak coupling'  really mean from the gravity side? 
    We want  the two  conserved spin-2 operators to be  replaced by one conserved operator and another
    that acquires a very small anomalous dimension.  Intuitively, this can happen not only if 
    the interaction Hamiltonian  is tunably small, but also if it involves a very small fraction of 
    the CFT   degrees of freedom. 
  With this in mind we distinguish two  holographic mechanisms 
  for coupling the two Universes: 
 \begin{itemize}
 \item   {Double-trace}   $\int \phi  \Phi$,   where $ \phi$ is an
 operator of cft$_d$ and   $\Phi$ an operator of  CFT$_d$; 
 
 \item Mediated by   {messengers},  such as in the gauging of a common global symmetry.
 
 \end{itemize}
  
   \noindent   To preserve  the AdS$_D$  isometries    these couplings must    be marginal,  or if they 
    are relevant  we follow them  to an  infrared fixed point. 
     Note that  since massive messenger fields
    can be integrated out   to give local multitrace couplings, the second mechanism  is distinct 
    from the first only if (some of)
    the messengers are massless.

      Let us  be a little   more precise about the meaning of  eq.\,\eqref{2spin2} in the second case,
      in which   
      the original Hilbert space   ${\cal H}_{\rm cft} \otimes {\cal H}_{\rm CFT}$ is typically  enlarged.
      If the messengers are scalar fields,   
       $t_{ab}$ and  $T_{ab}$ are still well defined in the new Hilbert space. 
     But if the messengers   mediate gauge interactions the above operators may  fail to be gauge invariant. 
        A simple fix is to remove from 
       the old $t_{ab}$ and $T_{ab}$ all the fields
       that participate in this gauge interaction. Our implicit assumption in what follows is that 
       the  linear combinations \eqref{2spin2} can be defined,  and are  close (in the sense
        of  operator norm)  to the two lowest-lying   spin-2 operators of the exact interacting  theory.

      Another  important remark is also   in order. In conventional holography one can only distinguish 
      single- from  multi-trace operators   in the   zero-string-coupling ($g_s$) limit. 
      At any finite  $g_s$  single-trace and multitrace
      operators  mix and eigenstates of the energy  are linear combinations
      of  both.  What we   mean, on the other hand,  by  `double-trace'  in this paper  are 
       products 
      of an operator of  cft$_d$ with one of  CFT$_d$. Such operators can be clearly distinguished 
     when  the two conformal theories don't talk, independently  of the
     string coupling constant.   A  more appropriate  name  would have been 
      `bridging' operators, but   we will keep refering to  them   as `double-trace.'

    
 \subsection{No multitrace for $N_Q >4$}\label{sec:3.2}
 
     Before discussing   other similarities or differences,  let us  see if  these  two mechanisms are
     compatible with  supersymmetry.  In section \ref{sec:2} we  ruled out,   on general grounds,   
          all cases   with $D>5$ and also all cases with $N_Q>8$.
       So  we need only consider  
       ${\cal N}\leq 2$ supersymmetries in AdS$_5$  and ${\cal N}\leq 4$ supersymmetries
        in AdS$_4$, i.e.  four-dimensional SCFTs  with ${\cal N}\leq 2$ and three-dimensional
       SCFTs  with ${\cal N}\leq 4$. 
            Now  gauging  a  non-anomalous global   symmetry  is always allowed  if  $N_Q\leq 8$  (the largest
        value   that admits 
      vector  multiplets)
       and the gauging  is   marginal or relevant
       in $d=4$ or $d=3$.  Thus  the `messenger  mechanism'  does not exclude
       more cases than those    already ruled out. 
     
             The `multitrace mechanism'  is  more  constrained. In this case the    deforming
              operators must    be top components of 
                 multiplets that arise in the  tensor product  of two  elementary ones, one from
                 each of the  two  decoupled theories, and their derivatives.
                    This condition  is impossible to  meet  while
preserving  $N_Q>4$ supercharges as I will now explain.

                 Consider 
             first   ${\cal N}= 4$,  $d=3$. There are   no    marginal  (Lorentz-invariant) 
             deformations in this case, and  the only  relevant ones   are  (i) 
             hypermultiplet masses and Fayet-Iliopoulos terms,  or  
             (ii) a universal
             mass deformation that resides in the multiplet of the conserved  energy-momentum
              tensor \cite{Cordova:2016xhm}.  None of these can couple two non-trivial  
              decoupled theories. 
         The standard mass deformations may  only couple two free hypermultiplets. Likewise
          Fayet-Iliopoulos terms can only couple 
         two free twisted hypermultiplets,\,\footnote{These statements do not rely on a Lagrangian description.
         In the notation of ref.\,\cite{Cordova:2016emh} the  deformations reside in $B_1[0]^{(2,0)}$
         multiplets (or in  their mirrors). The `multitrace'  property requires them to be  a tensor product 
         of  two $B_1[0]^{(1,0)}$ which are free hypermultiplets  QED. 
          } 
        while   the  universal mass deformation flows to a gapped theory and is for us  uninteresting.

               The    argument actually   works also  in the less supersymmetric   ${\cal N}= 3$, $d=3$ case. 
               There are again  no marginal deformations, and the  only relevant ones are flavour masses that
               can  only couple  two free hypermultiplets. Finally we  rule out 
                ${\cal N}= 2$  in $d=4$. Here the relevant and marginal deformations are of
                two kinds:  (i) [Higgs-branch]  flavour mass deformations, and (ii)  [Coulomb-branch]
                deformations that reside
                in chiral multiplets with $U(1)_R$ symmetry charge $2< r\leq 4$  \cite{Cordova:2016xhm, Argyres:2015ffa}. None of these    is
                a  product of   ${\cal N}= 2$ multiplets except in the trivial case of free
                 fields.\,\footnote{The statement  does not again rely on a Lagrangian description.
                The two types of operators reside in  multiplets (i) $B_1\bar B_1[0]^{(2;0)}$   
                    or  (ii) $L\bar B_1[0]^{(0;r)}$ where the superscript gives the $SU(2)_R\times U(1)_R$
                    quantum numbers,  and  $2< r\leq 4$  
                      \cite{Cordova:2016emh}.
                    These  operators can   only be  tensor products of  the multiplets  $B_1\bar B_1[0]^{(1;0)}$ or 
                    $A_2 \bar B_1 [0; 0]^{(0; 2)} $  which describe, respectively,  a free hypermultiplet
                    and a free vector multiplet.
                    } 
                The marginal Coulomb-branch operators,  in particular,  
                 are all believed to be   gauge-coupling and theta-angle deformations. I  will comment more on
                 these   in the following section.

             The cases ${\cal N}= 2$,  $d=3$ and ${\cal N}= 1$,  $d=4$ are not   ruled out as
               easily seen with explicit examples. The marginal or relevant deformations for  ${\cal N}= 1$,  $d=4$
             are superpotential deformations    in 
                 chiral multiplets with $U(1)_R$ charge $2\geq r > {2\over 3}$. One can form a marginal
                 deformation $(r=2)$ from the product of two   chiral multiplets with $r=1$. A famous example is
                 the Klebanov-Witten theory \cite{Klebanov:1998hh} in which the bifundamental chiral
                 fields have $R$ charge $r= {1\over 2}$, so there are   gauge-invariant operators with $r=1$. 
             The  product of two such operators can be used to deform  two previously decoupled  
             Klebanov-Witten    theories. 
             The case  ${\cal N}= 2$,  $d=3$ is even simpler, since quartic superpotentials  are classically marginal. 
             They  can be used to make   two decoupled theories interact. For instance ${\cal N}= 4$ SCFTs have 
                   a plethora 
             of   marginal  ${\cal N}= 2$    deformations of this kind (see e.g.\,\cite{Bachas:2017wva}).

                    These  conclusions are summarized in table \ref{tab:12}\,.  
                    Also listed  in this table are   the  
                    multiplets that  provide the  extra polarization states  of a massive supergraviton
                    in all the cases where Higgsing is a priori allowed. 
                    Consistency requires that such Stueckelberg 
                    multiplets should be available  whenever two 
                    decoupled theories are made to interact. We will now  see that  this is indeed 
                    always the case.

 \begin{table}[!h]
   \vskip 5mm
 \begin{center}
\begin{tabular}{ c |c | c |c |c |}
 \cline{2-5}   
 &  {\bf Susy} & \bf Multitrace  & \bf \,   Massless graviton  & \bf \,   Stueckelberg
  \\   [0.7ex]    \hline\hline 
  \multicolumn{1}{ |c |} {\multirow{2}{*}{\,\,\, AdS$_5$}  \,\,\, }  & 
\multicolumn{1}{ | c| } {  ${\cal N}$=2} & {\rm no}  & $ A_2\bar A_2[0;0]^{(0;0)}_2  $
& $B_1\bar B_1[0;0]^{(4;0)}_4  \oplus \bigl( A_2\bar B_1[0;0]^{(2;2)}_3 \oplus$ cc$ \bigr)$
\\    [0.5ex]  \cline{2-5}  
\multicolumn{1}{ |c| }{}    &
  \multicolumn{1}{ |c |} {${\cal N}$=1}   & {\rm yes} & $A_1\bar A_1[1;1]^{(0)}_3 $
  & $L\bar A_2[1;0]^{(1)}_{7/2} \oplus $ cc \\      [0.5ex]   \hline\hline 
   \multicolumn{1}{ |c |} {\multirow{4}{*}{\,\,\, AdS$_4$}  \,\,\, }  & 
\multicolumn{1}{ | c| } {  ${\cal N}=4$} & {\rm no}  & $ A_2[0]^{0, 0)}_1$   & $B_1[0]^{(2,2)}_2$  \\    [0.5ex]  \cline{2-5} 
\multicolumn{1}{ |c| }{}    &
  \multicolumn{1}{ |c |} {  ${\cal N}=3$}   & {\rm no} & 
  $A_1[1]^{(0)}_{3/2} $
  & $A_2[0]_2^{(2)}$  \\      [0.5ex] \cline{2-5}
  \multicolumn{1}{ |c| }{}    &
  \multicolumn{1}{ |c |}  {${\cal N}=2$}    & {\rm yes} & $A_1\bar A_1[2]_2^{(0)} $
  & $L\bar A_1 [1]_{5/2}^{(1)}\,\oplus$ cc  \\      [0.5ex] \cline{2-5}
  \multicolumn{1}{ |c| }{}    &
  \multicolumn{1}{ |c |}    {${\cal N}=1$}    & {\rm yes} & $A_1[3]_{5/2}$
  & $L[2]_{3}$  \\      [0.5ex] \hline\hline
 \end{tabular}  
   \medskip  
\caption{\small The  AdS supergravities for which  
the  graviton multiplet can obtain a mass by  gauging a global symmetry of the dual SCFT. 
The alternative `multi-trace mechanism' is only possible for $N_Q\leq 4$ supercharges. 
    The superconformal algebras  are  
$\mathfrak{su}(2,2\vert {\cal N})$ in $d=4$ and 
$\mathfrak{osp}( {\cal N}\vert 4)$ in $d=3$. The two right-most  columns list 
the massless graviton  and the Stueckelberg multiplets  
  in  the notation of ref.\,\cite{Cordova:2016emh}. The massive
graviton  multiplets
contain  $2^{N_Q-1}$ bosonic  and $2^{N_Q-1}$  fermionic   states. 
    }   \label{tab:12} 
 \end{center}
 \end{table}


 \subsection{Stueckelberg multiplets}\label{sec:3.3}

       Since the reader may not be familiar with   the  representation-theoretic notation of
      ref.\,\cite{Cordova:2016emh},  let me discuss the case of ${\cal N}=1$ supersymmetry in $d=4$
      in  more familiar  language.   In this case the 
       conserved and traceless energy-momentum tensor sits   in a real vector superfield  
       $R_{\alpha\dot\alpha}$,  subject to  the conditions
        $\bar D^{\dot\alpha}R_{\alpha\dot\alpha}=  D^{\alpha}R_{\alpha\dot\alpha} = 0$. 
         This superfield includes in addition to $T_{ab}$  the conserved 
         $R$-symmetry   current $R_a$   (see \cite{Derendinger:2016iwb} for a nice  discussion
         of the energy-momentum superfield for ${\cal N}=1$  in $d=4$).  
       In the notation of ref.\,\cite{Cordova:2016emh}   this is the multiplet $A_1\bar A_1[1;1]^{(0)}_3 $.  The    
       conditions  $\bar D^{\dot\alpha}R_{\alpha\dot\alpha}=  D^{\alpha}R_{\alpha\dot\alpha}=0$ 
        imply  indeed  that null states appear    at level 1 as indicated by the subscripts of $A$.

             Initially, each of the two  decoupled  SCFTs  has its own 
             conserved $R$ superfield. After coupling
         these $R$-superfields   mix linearly 
          like  their spin-2   components  in eq.\,\eqref{2spin2}, so that the  two linear
        combinations  now obey the
       conditions
       \bea
      {D^\alpha {\cal R}_{\alpha\dot\alpha}=} \bar D^{\dot\alpha}{\cal R}_{\alpha\dot\alpha}= 0\ , \qquad 
        \bar D^{\dot\alpha}{\cal R}^\prime_{\alpha\dot\alpha}=  V_\alpha \quad {\rm and} \quad
        D^{\alpha}{\cal R}^\prime_{\alpha\dot\alpha}=  - \bar V_{\dot \alpha} \ .  
       \eea
           Here ${\cal R}_{\alpha\dot\alpha}$ is the sum of  $R$  superfields  which remains  conserved,   and 
       $V_\alpha$ is the Stueckelberg superfield  which obeys the  second-order condition 
       $\bar D^2 V_\alpha =0$ and 
          renders
       (in the dual supergravity)  the second graviton massive. 
       The $L\bar A_2[1;0]^{(1)}$ representation of 
        $\mathfrak{su}(2,2\vert 1)$ in  table \ref{tab:12},
       with null states at   level 2, corresponds precisely to the multiplet $V_\alpha$. 
       Note that   the familiar  Ferrara-Zumino superfield
       \cite{Ferrara:1974pz}  is also defined as a pair ($R_{\alpha\dot\alpha}, V_\alpha$) 
       but with  
     $V_\alpha$   further
       constrained to ensure  energy-momentum  conservation \cite{Derendinger:2016iwb}. 
      In our case, after coupling the two theories, 
         none of the currents
        in  the superfield  ${\cal R}^\prime_{\alpha\dot\alpha}$  are  conserved. 
    \smallskip
            
         Consider   now  the marginal  superpotential   $W= \phi \Phi$ 
          with  $\phi$  a chiral  superfield  of cft$_4$,  $\Phi$ a chiral  
            superfield  of CFT$_4$, and   the sum of  $R$ charges equal  to 2.  It is easy to see that the
            spinor  superfield $V_\alpha:=  
            \phi D_\alpha \Phi -  \Phi D_\alpha \phi $ has the desired properties to be  
             the Stueckelberg field. 
       It has  the right  scaling dimension and $R$  charge, and  obeys the  second-order condition 
       $\bar D^2 V_\alpha =0$. 
       The story  generalizes to all   cases in  table \ref{tab:12} that admit   marginal double-trace deformations. 
       Let   the same symbol $\phi$ 
         denote  the superconformal representation that includes  (in the sense
       of the operator-state correspondence) the chiral superfield   
       and all of  its  derivatives. One finds
       in all $N_Q\leq 4$  cases the schematic decomposition                            \bea
         \phi  \otimes  \Phi  \   = \  {\rm deformation} \oplus  {\rm Stueckelberg} 
       \oplus \cdots 
       \eea
  showing that  when a marginal double-trace deformation is available 
  the Stueckelberg  field is automatically in place.\,\footnote{For  ${\cal N}=1$, $d=3$  
    the Stueckelberg superfield is a long multiplet, another manifestation of
  the fact that marginal deformations are in this case accidental. In the dual ${\cal N}=1$ AdS$_4$ supergravity
   this multiplet has six  bosonic degrees of freedom giving (with  the massless graviton)   a massive
   spin 2 and a massive spin 1.}
                                                                                                                                                                                                                                                                                                                   

                                                                                                                                                                                                                                                                                                                  What  about $N_Q>4$ and the messenger mechanism?  
                                                                                                                                                                                                                                                                                                                  Consider  ${\cal N}=2$ in $d=4$. The basic ingredients  for gauging a global symmetry are
                                                                                                                                                                                                                                                                                                                   conserved currents of  hypermultiplet global symmetries and free vector
                                                                                                                                                                                                                                                                                                                  multiplets. These  correspond, respectively,  to the two unitary representations   $B_1\bar B_1[0;0]_2^{(2;0)}$  and  $A_2\bar B_1[0;0]_1^{(0,2)}$  of $\mathfrak{su}(2,2\vert 2)$ \cite{Cordova:2016emh}.  Tensor products of these
                                                                                                                                                                                                                                                                                                                  representations give precisely the
                                                                                                                                                                                                                                                                                                                  Stueckelberg multiplets in table \ref{tab:12}\,.\footnote{The only other way to obtain  these Stueckelberg
                                                                                                                                                                                                                                                                                                                   multiplets
                                                                                                                                                                                                                                                                                                                  as tensor products  is from   $B_1\bar B_1[0;0]_1^{(1;0)}$ and $A_2\bar B_1[1;0]_2^{(1,2)}$
                                                                                                                                                                                                                                                                                                                  which correspond, respectively, to a free hypermultiplet and an extra conserved supercurrent multiplet \cite{Cordova:2016emh}.   Since the latter is not usually available in the decoupled SCFTs, 
                                                                                                                                                                                                                                                                                                                   gauging a global symmetry 
                                                                                                                                                                                                                                                                                                                  is  the only generic way to couple  weakly  two non-trivial 
                                                                                                                                                                                                                                                                                                                  $d=4$ SCFTs  while respecting ${\cal N}=2$ superconformal invariance. 
                                                                                                                                                                                                                                                                                                                  } 
                                                                                                                                                                                                                                                                                                                  
                                                                                                                                                                                                                                                                                                                  Things work out similarly for ${\cal N}=4$ in $d=3$. One  difference is that  a  free vector multiplet  
                                                                                                                                                                                                                                                                                                                  has an  associated  topological current  $\epsilon^{abc}F_{bc}$,  so the appropriate 
                                                                                                                                                                                                                                                                                                                   superconformal representation is that of 
                                                                                                                                                                                                                                                                                                                  a conserved-current multiplet $B_1[0]_1^{(0;2)}$. Its product with a [Higgs-branch] conserved current
                                                                                                                                                                                                                                                                                                                  $B_1[0]_1^{(2;0)}$ gives precisely the Stueckelberg superfield of  table \ref{tab:12}\,. This shows that the Stueckelberg fields for  Higgsing the supergraviton are automatically in place also for   the `gauging
                                                                                                                                                                                                                                                                                                                   mechanism'.

                                                                                                                                                                                                                                                                                                                   Note that the   Stueckelberg multiplet can be thought of intuitively   as the 
                                                                                                                                                                                                                                                                                                                   superspace  diffeomorphism 
                                                                                                                                                                                                                                                                                                                    needed to  map  one Universe to the other. 
                                                                                                                                                                                                                                                                                                                   

\section{On  massive AdS supergravities}\label{sec:4} 

 The  microscopic  embedding of ref.\,\cite{Bachas:2017rch, Bachas:2018zmb} and the discussion
 of the previous section suggest  the existence of effective theories of a massive supergraviton
  with  ${\cal N}\leq 2$
supersymmetries  in five dimensions,  and  with ${\cal N}\leq 4$  in four dimensions. 
In  the maximally-supersymmetric allowed cases 
the  massive supergraviton contains $128$ bosonic and $128$ fermionic fields,
which split into those of a massless multiplet  ($24$    bosons and as many fermions in $D=5$,  and
16 in $D=4$) 
and a Stueckelberg
multiplet (respectively $104$  and $112$ bosons and as many fermions).  A
possibly significant coincidence is  that the total number
of fields in these  massive-graviton  multiplets  
 is the same as for  the familar maximal (N=8)   supergravities in $D=4,5$ dimensions.\,\footnote{I thank Pierre Fayet for   this comment.} 
                                                                                                                                                                                                                                                                                                                   
        Constructing these effective massive supergravities is a nice technical  challenge. 
        Massive linearized actions for ${\cal N}=1$ in flat 
        $D=4$ dimensions   have been constructed 
           in refs.\,\cite{Zinoviev:2002xn, Gregoire:2004ic, Ondo:2016cdv} but little is known for extended
           supersymmetry and for AdS  backgrounds. In the case $D=4$,  ${\cal N}=4$ one could in 
           principle compute the effective action from the 
            string theory embedding  \cite{Bachas:2017rch, Bachas:2018zmb}, 
        but in practice 
         this is a formidable task. The  direct construction   is a more realistic
         project.   Here   I   assume that   
   the  putative supergravities  (for  all cases of  table \ref{tab:12}) do  exist, and make 
       some  comments  about their validity range.


\subsection{Breakdown  and a distance conjecture}\label{sec:31}

   The range of validity of the effective theories is a central issue 
  in the bottom-up approach to massive gravity. 
      The question is qualitatively different in the Minkowski and   Anti-de Sitter
   backgrounds because the scalar mode  of the graviton behaves differently
   in the two cases.  
       In the AdS$_D$  case of interest to us  here,  
 the largest energy scale  $ \Lambda_\star$ at which  all  effective theories are believed to 
        break  down  is
  \cite{deRham:2016plk} 
      \bea\label{bnd}
   \Lambda_\star^{D+2}  \sim   {m_g^2 m_{\rm Pl}^{D-2}\over l_{\rm AdS}^2}\ . 
   \eea
 This limit follows from a   power-counting argument for  
  the    optimal non-renormalizable
  action of  the Stueckelberg field (which comes entirely from
  the non-linear completion of the mass  term). 
    The argument  is
  robust  enough
  to [most likely]   survive supersymmetric completions. In flat spacetime it was indeed shown \cite{deRham:2018qqo, Bonifacio:2019mgk}
  that no finite number of spin $<$ 2 fields can push the
  corresponding  breakdown scale to   higher energy.

  One may convert this   $ \Lambda_\star$ scale
    to a  cutoff  on scaling dimensions in the conformal theory  by
 multiplying   both
 sides  of  \eqref{bnd} 
 with  $l_{\rm AdS}^{D+2}$.
 For  spin-2 fields  whose mass-dimension relation  is  $\Delta_* (\Delta_* - D+1) = \Lambda_\star^2\,  l_{\rm AdS}^2$\,,   this gives
   \bea\label{bnd2}
   \Delta_*  \sim  \begin{cases}
      & (D-1)  +  (\epsilon_g  c)^{2\over D+2}\quad {\rm if   } \  \epsilon_g c \ll 1 \ ,  \\ &
   (\epsilon_g  c)^{1\over D+2} \qquad \qquad {\rm if   } \  \epsilon_g c \gg  1 \ .
   \end{cases}
 \eea
 The meaning of the above estimate is as follows.  
 In any CFT  with holographic dual and  in  which the
   lowest spin-2 operator has 
 anomalous dimension $\epsilon_g \ll 1$,  
 new operators of spin $\geq 2$ will enter 
 at scaling dimensions  $\sim \Delta_\star$,  where  $c$ is  the central charge of the CFT. 
 
 \smallskip
       Actually only the upper branch of the bound   \eqref{bnd2}  is  relevant to our discussion
       here. 
 This is because towers of spin-2 excitations with scaling-dimension spacings  $\sim O(1)$ are
   anyway a  generic feature of  all  holographic models. Such towers 
   are inevitable  in theories  with $N_Q >  2$ supercharges which have continuous
      $R$  symmetries. The towers in this case  correspond to Kaluza-Klein excitations  
      along  the  compact submanifold whose 
  isometries  realize the continuous $R$ symmetry. {In addition  multitrace
  operators of spin-2 are   generic at $\Delta \sim O(1)$ and hard to disentangle
  from single-trace operators}.\,\footnote{This is sometimes refered to as  the AdS `scale separation
  problem'.  The problem could be in principle 
   avoided for   ${\cal N}=1$ in AdS$_4$, the  case  where marginal
  deformations need  fine-tuning and are   accidental. Even in this case, however,  one expects towers of
  higher-spin multi-particle states with spacings $\sim O(1)$, because of the attractive nature of
  gravity and the focussing effect of AdS spacetime.
 }   
 
 We assume then  that $\epsilon_g c \ll 1$, i.e. that the cutoff $\Lambda_\star$ is much below the
 characteristic AdS Hubble scale.    
      In this case the bound  \eqref{bnd2} predicts a condensation of spin-2 modes, so
      that  the limit
      of zero graviton mass 
       should be a decompactification limit at infinite distance. 
       One can  
           formulate this as a `distance conjecture'  for  massive  spin-2 particles in 
   AdS.\,\footnote{See 
   \cite{Klaewer:2018yxi, deRham:2018dqm, Palti:2019pca} 
   for recent discussions of  swampland conjectures involving spin-2 fields in  flat spacetime.
  }  \bigskip\smallskip

 \fbox{%
    \parbox{14.4cm}{%
         {\bf Massive-AdS-graviton conjecture}: \\ \vskip -3mm   If  the lowest-lying 
  graviton mass ($m_g$)  varies continuously in  a family of 
  AdS$_D$  vacua of string theory, 
  then    $m_g=  0$
  is at infinite distance in moduli space. Furthermore,   approaching   this point brings down a 
    tower of spin $\geq 2$  excitations with  mass spacings  vanishing   at least as  
  $  \Lambda_\star \sim m_g^{2/D+2}m_{\rm Pl}^{D-2/D+2}$ in units of the AdS radius, or faster. 
   }%
}            
    
\bigskip\smallskip

    \noindent  Both statements  can be 
      verified  in the microscopic  embedding  of ${\cal N}=4$  AdS$_4$ 
     massive gravity,   as I  will explain   in the next
     subsection. In the case of ${\cal N}=2$  AdS$_5$ the parameter controlling the graviton mass 
     is typically  
     a marginal gauge coupling, $\tau$,  of  the SCFT$_4$. The decoupling
      limit $\tau=\infty$ is  at infinite distance  in  simple examples, but it would be interesting to 
      give a general proof.  Furthermore, 
      characterizing  the decoupling   limit  more precisely should  allow to
       reinforce or invalidate the second part of
      the  conjecture.

   
  \subsection{Janus throat and the $\Lambda_*$ scale} 
   
       The string-theory embedding of refs.\,\cite{Bachas:2017rch, Bachas:2018zmb}  
         is dual to a pair of  ${\cal N}=4$, $d=3$ superconformal theories coupled by 
         gauging a  common `small'   flavour symmetry,   $U(n)$  with   $n^2 \ll c, C$. 
         Since  gauging is a relevant 
         deformation  in three dimensions,  we let    it  flow to an infrared 
         fixed point. This is in general a strongly-coupled fixed point, yet 
         the effect on the low-lying spin-2 spectrum can be arbitrarily small. 
          Both the disjoint theories  and the coupled one  are SCFTs 
                 of the type conjectured
         by Gaiotto and Witten \cite{Gaiotto:2008ak}.

          The corresponding type IIB  solutions  were derived for this class of theories in
           refs.\,\cite{Assel:2011xz, Assel:2012cj}.  In the limit of interest their key feature
           is a  Janus throat carrying $n$ units of five-form flux, and capped off at both sides
           by large compact six-manifolds  ${\cal M}_6$ and ${\cal M}_6^\prime$  (see the figure).  
          The (warped) compactification   AdS$_4\times_w{\cal M}_6$
           with AdS radius $l_{\rm AdS}$ is dual to  cft$_3$,  and  CFT$_3$ is 
           likewise dual to AdS$_4\times_w{\cal M}_6^\prime$
           with AdS radius $L_{\rm AdS}$. 
  One can think of the throat region as the hologram  of the
           $U(n)$ messenger degrees of freedom  connecting    the two theories.

        \begin{figure}[!htb]
\centering 
\includegraphics[width=.82\textwidth,scale=0.86,clip=true]{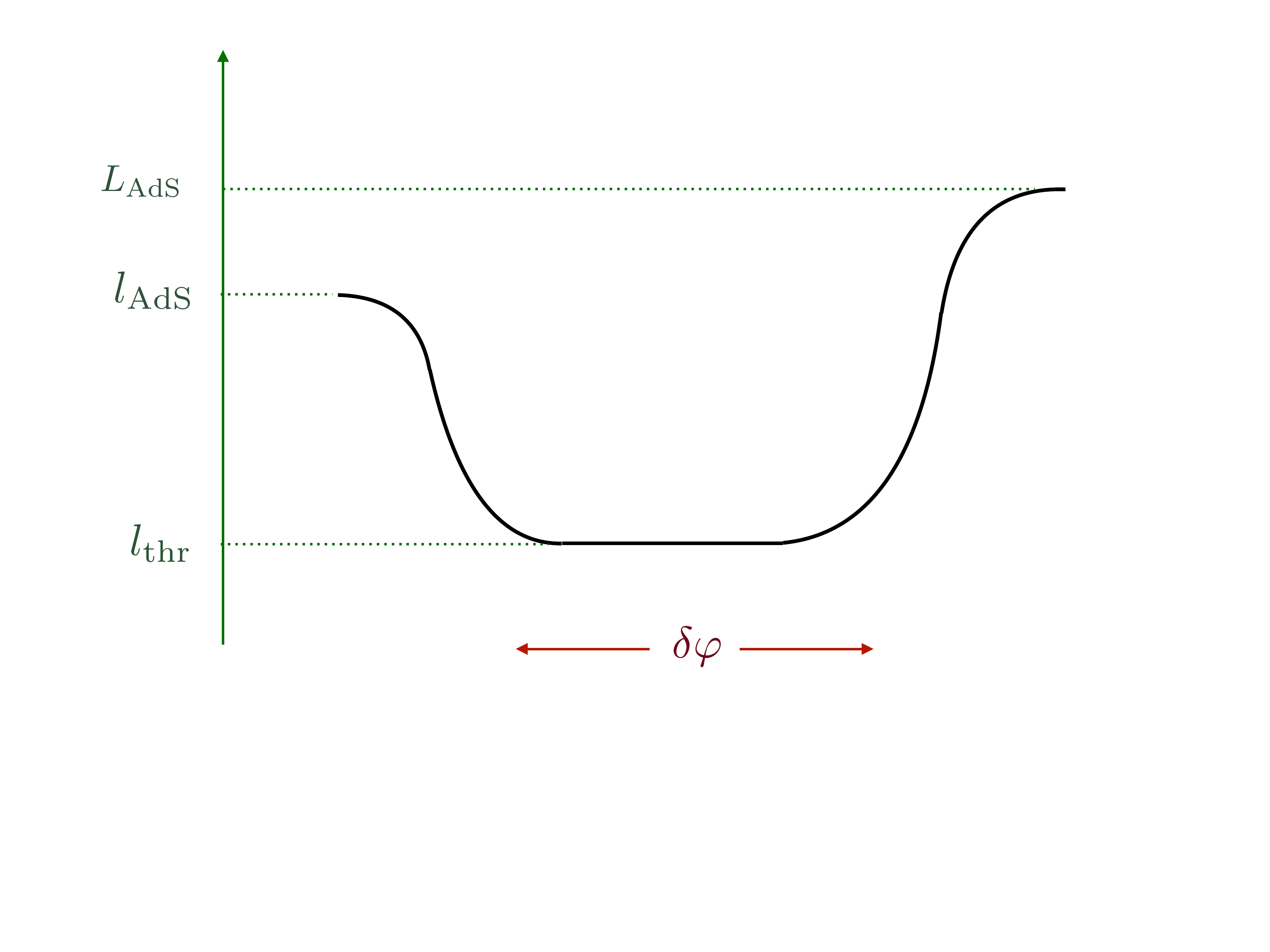}
    \vskip -19mm
   \caption{\footnotesize Schematic plot of the Janus metric
      $ds^2_{\rm Janus} \simeq l_{\rm thr}^2 dx^2 + \ell(x)^2 ds^2_{\rm AdS_4}$. 
    The horizontal and vertical axes parametrize, respectively,  
    the coordinate  $x$ and   the radius $\ell(x)$ of the AdS$_4$
    fiber.  The  relevant parameters are  the asymptotic left and right radii,  $l_{\rm AdS}$
    and $L_{\rm AdS}$,   the radius  $l_{\rm thr}$ of the throat, and 
   the size of its flat region $\delta x\sim 
   \delta\varphi$  over which the string-theory dilaton varies  linearly.
  }
\label{fig:3} 
 \end{figure}      
     
      The supersymmetric Janus solution was found in ref.\,\cite{DHoker:2006vfr}. 
      It extrapolates between  two asymptotic AdS$_5\times$S$^5$ regions  
      where the metric approaches\,  
      $ds^2 \simeq  l_{\rm thr}^2( dx^2 + \cosh^2x ds^2_{\rm AdS_4}$), 
      and the string-theory dilaton asymptotes to constant values $\varphi$ and $\varphi^\prime$. 
  The dilaton varies linearly  
   in a region of size $\delta x \simeq \delta\varphi  = \vert\varphi - \varphi^\prime\vert$ at the bottom of the throat. 
   For  $\delta\varphi = 0$ the solution reduces to  the standard AdS$_5\times$S$^5$   geometry
   with  radius $l_{\rm thr}  \sim n^{1/4}$. In our setup the two asymptotic regions are capped
   off  at some large values of $x$ on the left and the right. 
   One can think of these caps as `end of the world' branes in a 
   Randall-Sundrum  compactification  \cite{Randall:1999ee, Karch:2000ct}. 
   However, not only these branes are not thin,  but on the contrary  they
    occupy   most of spacetime. 
   
      {The full string theory solution has a large number of parameters  describing 
          the quivers of cft$_3$ and
          CFT$_3$,  but only five are relevant for the description of the Janus bridge: 
              the three radii  ($ l_{\rm thr}\ll  l_{\rm AdS}, L_{\rm AdS}$) and the two asymptotic values
          of the dilaton $\varphi$ and 
          $\varphi^\prime$.  Equivalently, these latter can be replaced by the effective four-dimensional
           Planck masses which, in units of the string scale, behave parametrically as 
          $$m^2_{\rm Planck} \sim l_{\rm AdS}^6 \, e^{2\varphi}\qquad
          {\rm and}\qquad M^2_{\rm Planck} \sim L_{\rm AdS}^6 \, e^{2\varphi^\prime}\ . 
          $$
          On the field theory side the  five parameters are $n, c, C$ and the two trilinear couplings
          of the cft and CFT energy-momentum tensors, 
          $\langle ttt\rangle$  and $\langle TTT\rangle$. In the specific realization of ref.\,\cite{Bachas:2018zmb}
          the dilaton parameters are  
        ratios of electric-to-magnetic flavour-group ranks in the two quiver
          theories, or equivalently   ratios  of D5-to NS5 brane charges in the IIB solutions.   
     We  dont need the precise relation  between $ \varphi$, $ \varphi^\prime$
     and the parameters of cft$_3$ and CFT$_3$ for our purposes here, only  that these are rational parameters 
     which  can be varied independently from all  the others. }

          Let us now  fix  the left theory (cft$_3$) on which our privileged observer lives, and vary
          the parameters of the right theory (CFT$_3$).  To keep the discussion simple,
           I  take rightaway the limit 
         $L_{\rm AdS}\to\infty$ so that the right-hand-side of 
         the solution asymptotes to AdS$_5/\mathbb{Z}_2\times$S$^5$. 
    This  solution  is   dual to a familiar setup, namely ${\cal N}=4$, $d=4$  super Yang-Mills 
      interacting with a strongly-coupled  
             theory,  cft$_3$, on the boundary of half-space. The $4d$ Yang-Mills theory has gauge group $U(n)$ 
             with $n^2 \ll c$, so most of the degrees of freedom
           live on the boundary, not in the bulk.\,\footnote{`Bulk'  and `boundary'  here refer to the 
           conformal field theory side of the duality. On  the  gravity side 
           they correspond to the AdS$_5/\mathbb{Z}_2$ spacetime and the thick-cap 
           AdS$_4$ brane.} Its  only other parameter  
            is the   
           four-dimensional Yang-Mills   coupling constant which depends on  the dilaton, 
           $g_{\rm YM}^2\sim e^{-\varphi^\prime}$.  
           
             One expects intuitively that the leaking-out of energy   from 
           cft$_3$ can be   suppressed
                   in  two different ways:  (a) by the scarcity of bulk  degrees of
           freedom ($n^2\ll c$),  and (b) in the decoupling  limit  $g_{\rm YM}\to 0$, i.e. 
             $\varphi^\prime \to \pm \infty$ (the two signs are related by S-duality). Both factors
             should contribute to make the mass of the graviton parametrically small, as  is indeed
             confirmed by the calculation of  ref.\,\cite{Bachas:2018zmb}.

               The basic idea behind this calculation is that the wavefunction of the lowest-lying
               spin-2 mode  vanishes at $x\to\infty$ and approaches exponentially
            {fast  a value,   $\psi_0$,  at the other end of the Janus throat. 
               Here $\psi_0$ is the constant wavefunction of the massless graviton in the 
               AdS$_4\times_w{\cal M}_6$ vacuum dual to the boundary cft$_3$.}
                This mode  is a non-normalizable mode in the full Janus
               background, but it  becomes normalizable thanks to the  cutoff 
               in the $x\to -\infty$ region.  Calculating the  mass gives \cite{Bachas:2018zmb}
          \bea\epsilon_g \,\simeq \,    \underbrace{{ \,   n^2   \over 4\pi^4  }
                            { \tanh^3\delta\varphi \over [\delta\varphi - \tanh\delta\varphi]}}_{\lambda^2_{\rm eff}}\,
                            \times  { 1\over c  }  \ .  
\eea   
    This is indeed proportional    both to   ${n^2/c}$,  and to $\delta\varphi^{-1}$ 
    for large $\delta\varphi  $, i.e. to the two
    suppression factors  advertized above.          
\smallskip
    
    We may now  return  to the question of the breakdown scale $\Lambda_\star$. Let us   
  write $\epsilon_g   := \lambda_{\rm eff}^2/c$ by analogy with the result 
     of  conformal perturbation
    theory,  eq.\,\eqref{ahetal}, 
    for  CFTs coupled by a  double-trace deformation  \cite{Aharony:2006hz}. 
    We are interested in the limit $\epsilon_g \to 0$ in which the graviton becomes  massless.
    We could try to take this limit in two ways:
   \begin{itemize}
   \item  By taking $n\to 0$. This is possible  in supergravity where $n\sim l_{\rm thr}^4$ is a continuous
   parameter,  but in string theory $n$ [the D3-brane charge of the throat] is   quantized. 
   
   \item By taking $\delta\varphi\to \infty$. This is at infinite distance in moduli space,
   in agreement with the first part of our conjecture.  
    In this limit the Janus throat develops a 
   flat  region of invariant length  $\sim l_{\rm thr} \delta\varphi$, bringing down 
   spin-2 Kaluza-Klein excitations  with  mass  spacings  
   $\sim  (l_{\rm thr} \delta\varphi)^{-1}$\,.\,\footnote{The spin-2 Kaluza-Klein spectrum in the Janus background can be computed analytically  \cite{Bachas:2011xa}.
   The result   confirms 
   the above heuristic argument.} The dual tower 
    of spin-2 operators on the field theory side would thus have anomalous scaling 
   dimensions $\sim  (\delta\varphi)^{-2} \ll  (\delta\varphi)^{-1/3} \sim \Delta_*-3$,
   in agreement with  the second part of the massive-graviton conjecture.
   
    \end{itemize} 
    
      It is  striking   that the full string-theory embedding of the `bridge'  is instrumental for
      reaching these conclusions.  In ten-dimensional supergravity  the limit of vanishing  Janus
       radius   is smooth,  leaving behind harmless  coordinate singularities \cite{Assel:2011xz}\cite{Aharony:2011yc}. 
      Without the string theoretic  quantization of $l_{\rm thr}$ 
       this  would be in tension with the  breakdown of   
       the effective massive-gravity
       theory. From the CFT side on the other hand, although it might be sometimes clear  that the decoupling
       limit   is at infinite distance,  the nature of this  limit can be  obscure.  It is important in this regard to stress
       that,  in the above  example,  the condensing  spin-2 states on the gravity side 
        are in long unprotected multiplets. Even though the entire setup has ${\cal N}=4$ supersymmetry,
        {\it   the  singularity} of the effective gravitational theory 
         {\it has its origin in the non-BPS sector.}

   \smallskip
   
         These results raise two immediate questions.  First, can  the above   analysis   be extended
          to  five-dimensional ${\cal N}=2$ supergravity,   the other 
          maximally-supersymmetric case among the
         possibilities of  table \ref{tab:12}\,?
       The   
         menagerie of four-dimensional  ${\cal N}=2$  SCFTs  has been  greatly
         extended by the construction of  class-S   theories \cite{Gaiotto:2009we}, 
             and much    is known  
         about their moduli spaces  and  their supergravity duals \cite{Gaiotto:2009gz}. An analysis
         similar to the one presented here for ${\cal N}=4$  SCFT$_3$, 
          is possibly within reach.\,\footnote{C.B and  A. Tomasiello, work in progress. The
          superconformal index, a  powerful
          tool for the study of  ${\cal N}=2$  SCFTs,  is unfortunately not adapted to our problem. The index 
           is blind to long multiplets such as the conjectured 
          unprotected  towers of spin-2 states. Interestingly  
            the authors of 
        ref.\,\cite{Distler:2017xba} 
          extracted   the number of conserved
         energy-momentum tensors from the superconformal index of selected class-S theories.  
          This is a good  diagnostic for  the number of non-interacting SCFTs, but 
        not of the conjectured   singular behavior  under small marginal   interactions.
         }
              
  \smallskip
  
   The second question concerns double-trace deformations.   The  
   standard  recipe
   maps   them to  modified boundary conditions on the gravity side
    \cite{Witten:2001ua, Berkooz:2002ug}. But  this is a recipe 
     in  the   effective $D$-dimensional  supergravity and its   lift to ten dimensions, let alone
     its  embedding in the full string theory,  is problematic. The lesson from 
      our discussion in this paper   is that the  string-theoretic description of the 
      `double-trace bridge'  may be much richer than hitherto imagined. 
   
         It is logically conceivable  that the
      marginal double-trace modulus, $\lambda$,  of the CFT  is quantized, but 
        there is no such  precedent   in string theory that I am aware of.  
          The more plausible hypothesis is that, as was the case 
     for the Janus
     throat, the $\lambda\to 0$ limit  is at infinite distance
     and  brings down a tower of light spin $\geq 2$ states. 
   It might be possible   to study this question with the conformal bootstrap.

  \bigskip
   
{\bf Acknowledgements}: { 
I have benefited from discussions with   
  Ali Chamseddine,  Jean-Pierre Derendinger, Sergei Dubovsky, 
Marc Henneaux,  Chris Hull, Ioannis Lavdas, Bruno Le Floch,  George Papadopoulos, 
Massimo Porrati, Jan Troost, Alessandro Tomasiello and  Misha Vassiliev. 
 I am particularly indebted to   Claudia de Rham and Andrew Tolley for a discussion
 that motivated the analysis in section \ref{sec:4}\,. 
  I am also grateful to  the String Theory group at Imperial College for hospitality during
 part of this  work. }



\end{document}